%% file: main.tex
\documentclass[runningheads]{llncs}
\usepackage{graphicx}
\usepackage[noadjust]{cite}
\usepackage{amsmath,amssymb,amsfonts}
\usepackage{algorithmic}
\usepackage{textcomp}
\usepackage{array}
\usepackage{float}
\usepackage{gensymb}
\usepackage[inline]{enumitem}
\usepackage{hyperref}
\usepackage{filecontents, pgfplots}
\pgfplotsset{width=10cm,compat=1.9}
\usepgfplotslibrary{external}
\usepackage{tcolorbox}
\usepackage{subcaption}
\usepackage{cleveref}
\usepackage{mathtools}
\usepackage{listings}
\usepackage{fancyhdr}
\usepackage[bottom]{footmisc}
\begin{document}
\title{Training Multiscale-CNN for Large Microscopy Image Classification in One Hour}
%
%
%
\author{Kushal Datta\inst{*,1}\orcidID{0000-0003-1608-6040} \and 
Imtiaz Hossain\inst{*,2}\orcidID{0000-0001-6747-5906} \and
Sun Choi\inst{1}\orcidID{0000-0003-4276-7560} \and 
Vikram Saletore\inst{1}\orcidID{0000-0001-8642-539X} \and
Kyle Ambert\inst{1}\orcidID{0000-0002-1688-4408} \and
William J. Godinez\inst{3}\orcidID{0000-0003-4753-4942} \and
Xian Zhang\inst{2}\orcidID{0000-0002-7337-747X}}
%
%
\institute{Artificial Intelligence Products Group, Intel Corporation, USA 
\and
Novartis Institutes for Biomedical Research, Basel, Switzerland\\
\and
Novartis Institutes for Biomedical Research, Emeryville, CA, USA\\
\email{\{kushal.datta,sun.choi,vikram.a.saletore,kyle.h.ambert\}@intel.com}\\
\email{\{imtiaz.hossain,william\_jose.godinez\_navarro,xian-1.zhang\}@novartis.com}}

\maketitle              

	\begin{abstract} 
		Existing approaches to train neural networks that use large images require to either crop or down-sample data during pre-processing, use small batch sizes, or split the model across devices mainly due to the prohibitively limited memory capacity available on GPUs and emerging accelerators. These techniques often lead to longer time to convergence or time to train (TTT), and in some cases, lower model accuracy. CPUs, on the other hand, can leverage significant amounts of memory. While much work has been done on parallelizing neural network training on multiple CPUs, little attention has been given to tune neural network training with large images on CPUs. In this work, we train a multi-scale convolutional neural network (M-CNN) to classify large biomedical images for high content screening in one hour. The ability to leverage large memory capacity on CPUs enables us to scale to larger batch sizes without having to crop or down-sample the input images. In conjunction with large batch sizes, we find a generalized methodology of linearly scaling of learning rate and train M-CNN to state-of-the-art (SOTA) accuracy of 99\% within one hour. We achieve fast time to convergence using 128 two socket Intel® Xeon® 6148 processor nodes with 192GB DDR4 memory connected with 100Gbps Intel® Omnipath architecture.
		
    \let\thefootnote\relax\footnotetext{\linebreak\emph{* Published in International SuperComputing High Performance 2019 Workshops. \linebreak * Authors made equal contributions to the paper.}}
	\end{abstract}


	\input{introduction.tex}

	\input{mcnn_model.tex}
	\input{training.tex}

	\input{dataset.tex}

\input{data.tex}
	\input{performance}
	\input{discussion}


		
	\bibliographystyle{IEEEtran}
	\bibliography{bibliography}
\end{document}

%% file: introduction.tex
\begin{section}{Introduction}
	\label{sec:intro}	
	\noindent Biomedical image analysis has been a natural area of application for deep convolutional neural networks (CNNs). Several uses of CNN-related topologies have been proposed in radiology\cite{Arbabshirani2018, Akkus2017}, histopathology \cite{ciresan2013, Litjens2016, Janowczyk2016} and microscopy \cite{Kraus2017, Sommer2017, Ciresan2012} (for a review, see \cite{Litjens2017}). High-content screening (HCS) \cite{Mattiazzi2016, Boutros2015, Singh2014, Scheeder2018, zock2009, assay2014}, the use of microscopy at scale in cellular experiments, in particular, has seen progress in applying CNN-based analysis \cite{Kraus2017, Godinez2017, Godinez2018, Sommer2017, Ando2017}. Instead of the conventional analysis approaches where cellular objects are first segmented and then pre-defined features representing their phenotypes (characteristic image content corresponding to the underlying experimental conditions) are measured, deep learning approaches offer the promise to capture relevant features and phenotypes without \textit{a priori} knowledge or significant manual parameter tuning. In deep CNNs, the deeper layers pick up high-levels of organization based on the input of many features captured in previous layers. Typically, a pooling operation (or a higher stride length in the convolution filter) is used to subsample interesting activations from one layer to the next, resulting in ever-coarser ``higher-level'' representations of the image content.\\\\
	\noindent Despite the potential of deep learning in analyzing biomedical images, two outstanding challenges, namely the complexity of the biological imaging phenotypes and the difficulty in acquiring large biological sample sizes, have hindered broader adoption in this domain. To circumvent these challenges, architectural changes have been introduced into some models to make training easier without trading off model accuracy. One novel approach is to use wide networks, which explicitly model various levels of coarseness. In these topologies, several copies of the input image are downsampled and used to train separate, parallel convolutional layers, which are eventually concatenated together to form a single feature vector that is passed on to fully-connected layers (e.g., see Buyssens et al.\cite{Buyssens2012}). A recent application of this idea to HCS is the Multiscale Convolutional Neural Network (M-CNN) architecture \cite{Godinez2017}, which has been shown to be generally applicable to multiple microscopy datasets, in particular for identifying the effect of compound treatment. \\\\
\noindent The computational footprint of M-CNN, although relatively small as compared with other deep CNNs (e.g., Residual Neural Network 152), is still large when applied to high-content cellular imaging. Thus, it is important that model-related aspects of memory utilization and training performance are thoroughly understood, and that an end user knows \emph{a priori} how to get maximum performance on their hardware. Commercial cloud service providers (CSPs) like Microsoft, Google, or Amazon--as well as on-premise HPC centers in academia and industry--are exploring custom hardware accelerator architectures, such as application-specific integrated circuits (ASICs) \cite{jouppi2017} or GPUs, to expedite training neural network models. In spite of the popularity of these technologies, several factors such as higher financial cost of ownership, lack of virtualization and lack of support for multi-tenancy, leading to poor hardware utilization, may be cited as reasons to consider CPU-centric performance optimizations in reducing the time-to-train for such models. Importantly, since almost all data centers, are already equipped with thousands of general-purpose CPUs, it makes a strong case for such an approach. \\\\
\noindent Existing approaches to improve the time to train convolutional image classification neural network model such as M-CNN designed to work with large high-content cellular images have needed to either crop or down-sample the images during pre-processing. Other ideas are to restrict to small batch sizes or split the model across multiple devices due to the limited memory capacity available on GPUs or accelerator cards. However, these techniques can lead to longer time to convergence or time to train (TTT), and in some cases, lower model accuracy. CPUs, on the other hand, can leverage large memory. Our primary contributions include,

\begin{enumerate}
	\item Train M-CNN to achieve SOTA accuracy of 99\% on multiple CPU servers without tiling or cropping of input images or splitting the model
	\item Use large batch sizes per CPU exploiting large memory
	\item Use multiple training instances/workers per CPU node to improve utilization
	\item Use large batches and learning rate scaling to achieve fast convergence
\end{enumerate}

\noindent The ability to leverage large memory capacity on CPUs enables us to scale to larger batch sizes without having to crop or down-sample the input images. In conjunction with large batch sizes, we linearly scale learning rate with global batch size and train M-CNN to SOTA accuracy within one hour. We achieve this fast time to convergence using 128 two socket Intel® Xeon® 6148 processor nodes with 192GB DDR4 memory connected with 100Gbps Intel® Omnipath architecture.

	\begin{figure}[t]
	\centering
	\includegraphics[width=4.75in]{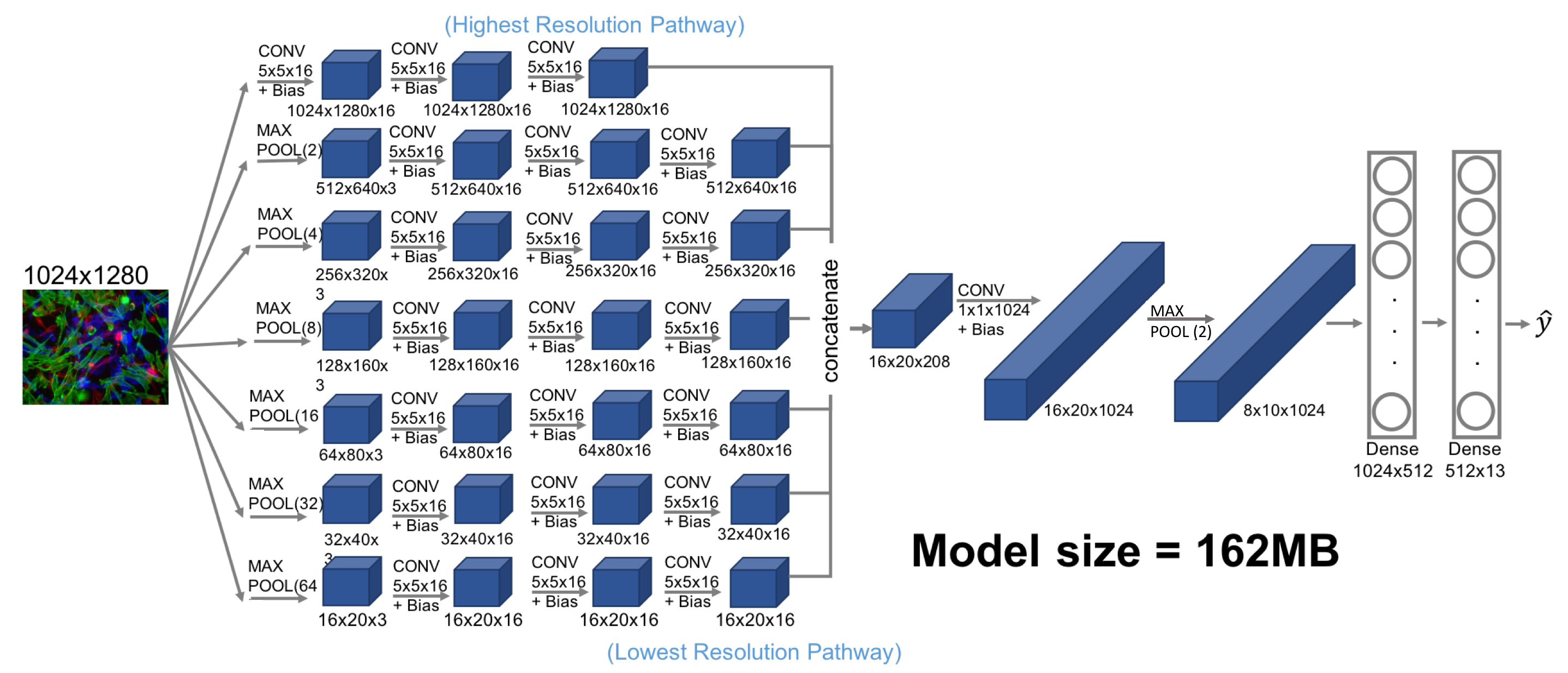}
	\caption{\textsf{Operations and kernels of the M-CNN model. Convolution is abbreviated \emph{CONV}, and Max Pooling operations are abbreviated as \emph{MAX POOL}}}
	\label{fig:mcnn}
\end{figure}

\end{section}

%% file: mcnn_model.tex
\begin{section}{Multi-scale convolutional neural network}
	\label{sec:mcnn}
	M-CNNs capture both fine-grained cell-level features and coarse-grained features observable at the population level by using seven parallel convolution pathways (\autoref{fig:mcnn}). As in \cite{Godinez2017}, image height and width are down-sampled by 64, 32, 16, 8, 4, and 2 times in the lower six pathways in ascending order, respectively, while images processed by the top-most path are operated on at the full resolution. The output of the last layers of convolution are down sampled to the lowest resolution and concatenated into a $16 \times 20 \times 208$ tensor. The concatenated signals are passed through a convolution with rectified linear activation (ReLU) and two fully connected layers. A final softmax layer transforms probabilistic per-class predictions associated with each image into a hard class prediction. In \autoref{fig:mcnn}, the size of convolution kernels are specified below the solid colored cubes, which represent the activations. The sum of the sizes of the convolution kernels and two dense layer, which are $1024 \times 512$ and $512 \times 13$, respectively, is 162.2 megabytes. Weights are represented as 32-bit floating point numbers.\\\\
	\noindent The network's gradient and activation size determine the lower bound of its memory footprint. We plot the calculated activation size of the feed forward network as the global batch size is scaled from 8 to 64 by factors of two in \autoref{fig:batchsize}. Note that the size of variables required for back propagation is identical to the size of the gradients and hence is determined by model size, not activation size.

	\begin{figure}[t]
			\centering
			\includegraphics[width=2.75in]{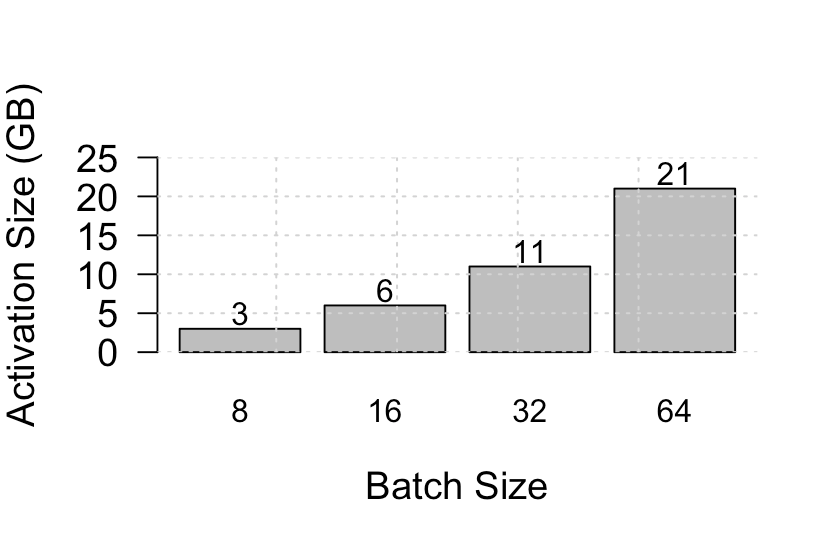}
			\caption{\textsf{Activation sizes in M-CNN as a function of batch size.}}
			\label{fig:batchsize}
	\end{figure}
	
\end{section}

%% file: training.tex
\begin{section}{Large batch training}
	\label{sec:training}
	Synchronous gradient descent and data-level parallelism are fundamental concepts to training a deep neural network. In this domain, the most common algorithm used for training is stochastic gradient descent (SGD), which exploits the fact that activation functions in a neural network are differentiable with respect to their weights. During training, batches of data are run through the network. This process is referred to as \emph{forward propagation}.  A loss function $E$ is computed at each training iteration, which quantifies how accurately the network was able to classify the input. The SGD algorithm then computes the gradient $\nabla_{W}(E)$ of the loss function with respect to the current weights $W$. On the basis of the gradients, weights are updated according equation \ref{eq:sgd}, where $W_{t+1}$ are the updated weights, $W_{t}$ are the weights prior to the adjustment (or previous iteration), and $\lambda$ is a tunable parameter called the learning rate (LR).
	\begin{equation}
		W_{t+1} = W_{t} - \lambda \nabla_{W} E
		\label{eq:sgd}
	\end{equation}

	\noindent Since each neural network layer is a differentiable function of the layer preceding it, gradients are computed layer-by-layer, moving from output to input in a process called backpropagation. Finally, the weights in the network are updated according to the computed gradient, and both forward and backpropagation are  repeated with a new batch of data. We continue repeating these procedures until the network has reached a satisfactory degree of accuracy on a hold-out validation data set. Training can require running millions of iterations of this process on a given dataset. The most popular approach to speeding up network training makes use of a data-parallel algorithm called synchronous SGD \cite{Robbins1951}. Synchronous SGD works by replicating SGD across compute nodes, each working on different batches of training data simultaneously. We refer to these replicas as \textit{workers}. A key requirement for synchronous SGD is for information to be synchronized and aggregated across all computing instances at each iteration. The update equation is show in equation \ref{eq:syncsgd}, where $B$ denotes the batch sampled from the training data, $n$ is the size of the batch.

	\begin{equation}
		W_{t+1} = W_{t} - \lambda \frac{1}{n} \sum_{x \in B} \nabla_{W}E(x)
		\label{eq:syncsgd}
	\end{equation}

	With $k$ workers each training with $B$ batches and learning rate $\lambda'$, we updates the weights according to

	\begin{equation}
		W_{t+1} = W_{t} - \lambda' \frac{1}{kn} \sum_{j < k} \sum_{x \in B_j} \nabla_{W}E(x)
		\label{eq:syncsgdmw}
	\end{equation}

	Thus, if we adjust the learning rate by $k$, the weight update equation stays consistent with the synchronous SGD update rule, helping the model to converge without changing the hyper-parameters. We refer to $n$ or $|B|$ as the \textit{local batch size}, and $kn$ as the \textit{global batch size}.

	\begin{subsection}{Learning rate schedule}
		\label{sec:lr}
		\noindent In addition to scaling the model's learning rate parameter (LR) with respect to the batch size, others \cite{You2017} have observed that gradually increasing it during initial epochs, and subsequently decaying it helps to the model to converge faster. This implies that LR is changed between training iterations, depending on the number of workers, the model, and dataset. We follow the same methodology. We start to train with LR initialized to a low value of $\lambda = 0.001$. In the first few epochs, it is gradually increased to the scaled value of $k\lambda$ and then adjusted following a polynomial decay, with momentum SGD (momentum=0.9).\\\\
\noindent Reaching network convergence during training is not guaranteed--the process is sensitive to LR values and features in the data. Scaling this process out to large batch sizes on multiple workers concurrently has the same considerations. If the per-iteration batch size is too large, fewer updates per epoch are required (since an epoch is, by definition, a complete pass through the training data set), which can either result in the model diverging, or it requiring additional epochs to converge (relative to the non-distributed case), defeating the purpose of scaling to large batch sizes. Thus, demonstrating scaled-out performance with large batches without first demonstrating convergence is meaningless. Instead, we measure the time needed to reach state of the art accuracy or TTT. The ingestion method for each worker ensures that each minibatch contains randomly-shuffled data from the different classes.
	\end{subsection}
\end{section}

%% file: dataset.tex
\begin{section}{Dataset}
	\label{sec:dataset}
	The Broad Bioimage Benchmark Collection BBBC021 image set \cite{bbbc021} is a collection of 13,200 images from compound treatment on MCF-7 breast cancer cells. Each image consists of three channels: the cells are labeled for DNA, F-actin, and B-tubulin and imaged with fluorescence microscopy. Metadata on compound treatment and concentration is also available\cite{Caie1913}. In all, 113 compounds have been used, each with varying concentrations and tested between 2 and 3 times each. Mechanism of action (MoA) labels are available for 103 compound-concentrations (38 compounds tested at between one and seven different concentrations each). In all, 13 MoAs (including the neutral control, DMSO) were available: 6 of the 12 MoAs were assigned visually.  DMSO treatments were treated as neutral control and assigned a separate label. The others were defined based on information on the respective compounds in the available literature.  We choose 1684 images from the BBBC021 dataset that are representative of all of the MoAs present. The distribution of the images according to MoA classes is shown in \autoref{fig:classdistribution}. The images are preprocessed and normalized as described in \cite{Godinez2017}. From the 1684 images, we create two datasets with different augmentation strategies:

		\begin{figure}[t]
			\centering
			\includegraphics[width=4.75in]{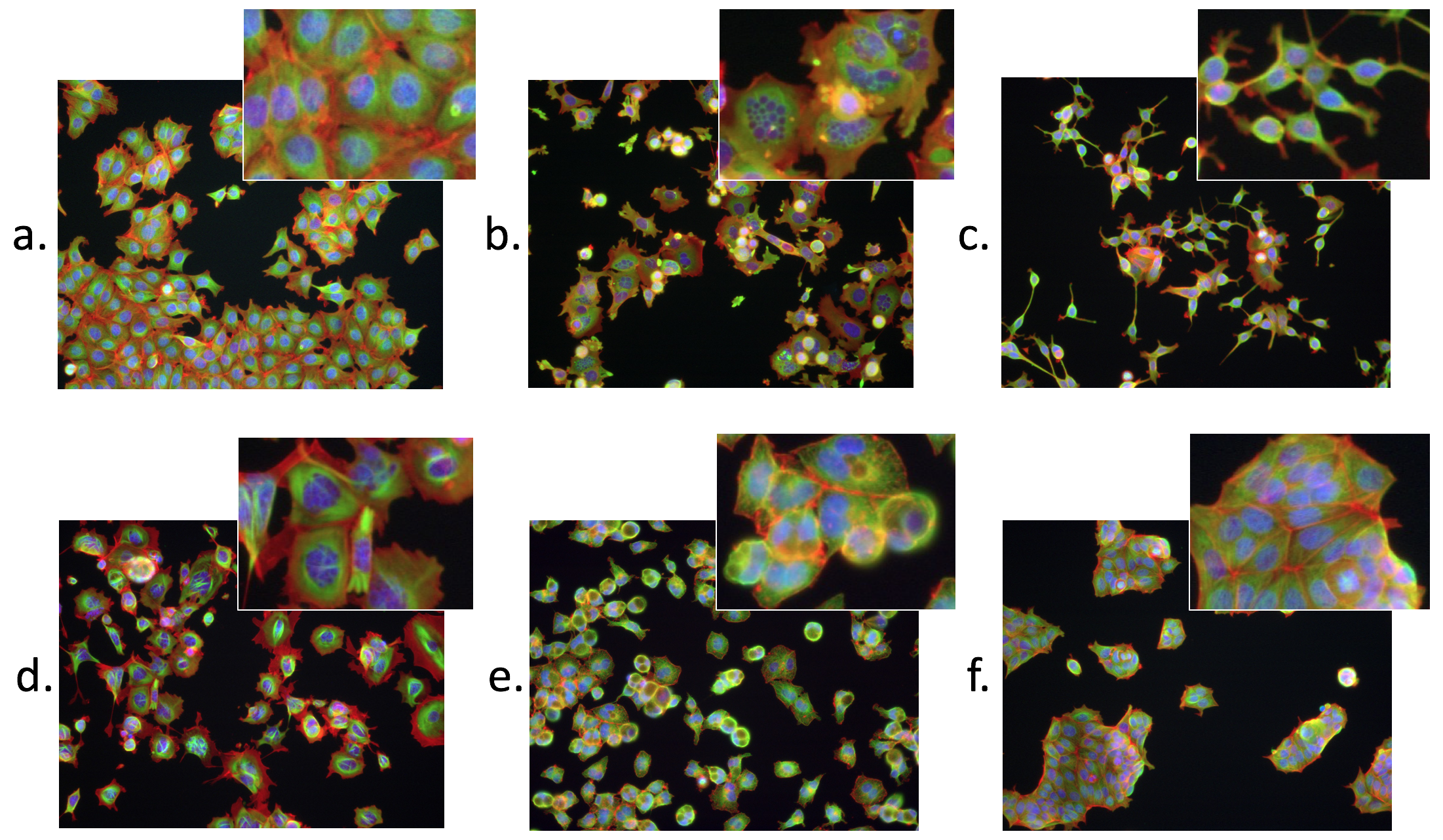}
			\caption{\textsf{Example images from the BBBC021\cite{bbbc021} dataset showing phenotypes from treatment with compound-concentration pairs with different mechanisms of action: a) DSMO (neutral control), b) Microtubule destabilizer, c) Cholesterol lowering, d) Microtubule stabilizer,  e) Actin disrupter, f) Epithelial. DNA staining is shown in blue, the F-actin staining in red and the Β-tubulin staining in green. The insets show a magnified view of the same phenotypes.}}
			\label{fig:bbbc_whole}
	\end{figure}
	
	\begin{itemize}
		\item \textit{Dataset A}: Images in this dataset are 1024x1280 pixesl wide with 3 channels. They are augmented to produce five copies as
		\begin{enumerate*}
			\item $90^{\circ}$ rotation,
			\item a horizontal mirror,
			\item vertical mirror,
			\item $90^{\circ}$ rotation of horizontal mirror and
			\item $90^{\circ}$ rotation of vertical mirror
		\end{enumerate*}. Total number of images in the dataset is $1684*6$ (five rotations + original) $= 10104$. We take a 90-10 split and create a training set of 9093 images and validation set of 1011 images. The total size of the images on disk are 38GB. 
		\item
		\textit{Dataset B}: This is a larger dataset. The dimensions of the images in this dataset are 724x724 pixesl with 3 channels. Similar to Dataset A, all images have 5 additional augmentations. Additionally, each image is rotated by $15^{\circ}$ to create 23 more augmentations. The total size of the images on disk are  512GB. Among them, 313282 images are used for training and 35306 are used for validation. \\\\
\noindent Ideally, we would have allocated a representative out-of-sample set of images as a validation set. However due to the paucity of MOA annotations in this dataset, and the fact that the main objective of this exercise is to reduce time to convergence, we allow for the fact that the validation dataset may contain an augmented version of an image in the training data, although never a copy of the same image. 
	\end{itemize}
	\begin{figure}[h]
	  \centering	
	    \includegraphics[width=3.5in]
	    {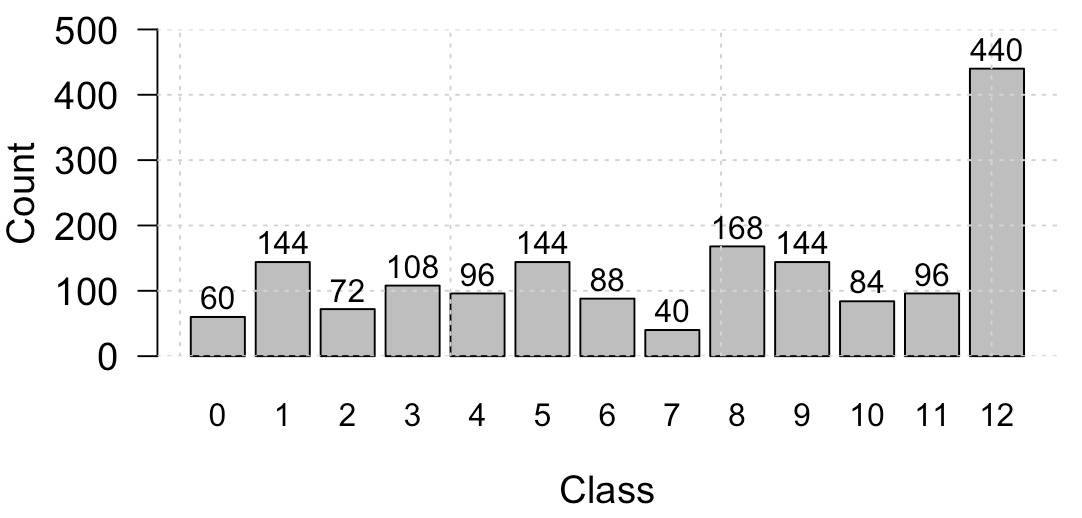}
	  \caption{Class distribution for the 1684 training images used in our experiment.}
	  \label{fig:classdistribution}
	\end{figure}
\end{section}

%% file: performance.tex
\begin{section}{Performance results}
	\label{sec:performance}
	\begin{subsection}{Experimental setup}
		\label{sec:setup}
		All experiments are run on two socket (2S) 2.40GHz Intel\textregistered{} Xeon\textregistered{} Gold 6148 processors. There are 20 cores per socket with 2-way hardware multi-threading. On-chip L1 data cache is 32KB. L2 and L3 caches are 1MB and 28MB respectively. For multi-node experiments, we used up to 64 Intel\textregistered{} Xeon\textregistered{} Gold connected via 100gigabits/second Intel\textregistered{} OP Fabric. Each server has 192GB physical memory and a 1.6TB Intel SSD storage drive. The M-CNN topology was added to the standard benchmarking scripts \cite{GoogleTPU} to leverage instantiation mechanisms of distributed workers. Gradient synchronization between the workers was done using Horovod, an MPI-based communication library for deep learning training \cite{Horovod}. In our experiments, we used TensorFlow 1.9.0, Horovod 0.13.4, Python 2.7.5 and OpenMPI 3.0.0.
	\end{subsection}
	\begin{subsection}{Scaling up TTT in One Node with Dataset A}
		\label{sec:scaleup}
\noindent We first performed a sweep of batch sizes from 4, 8, 16, 32 and 64 to check how fast we can converge on one CPU server. We acheived convergence in 5hrs 31mins with batch size = 32. The resulting throughput and memory consumed are shown in Figure \ref{fig:multiworker} \subref{fig:singlenodethroughput1w} and Figure \ref{fig:multiworker} \subref{fig:singlenodememutil1w}, respectively. As shown in the latter figure, the memory footprint of M-CNN far exceeds the activation size of the model. For example, in case of batch size of 32, total memory used is 47.5GB which is 4x larger than activation size of 11GB as calculated in \autoref{fig:batchsize}. The additional memory is allocated by TensorFlow to instantiate temporary variables used in both forward and backward propagation, buffers to read data and others operations. Due to these overheads, memory utilization of M-CNN is prohibitively high and it is difficult to scale to large batch sizes when memory in the system is limited. \\
		\begin{figure}[t]
	\centering
	\begin{subfigure}[t]{0.45\textwidth}
		\centering
		\includegraphics[width=\textwidth]{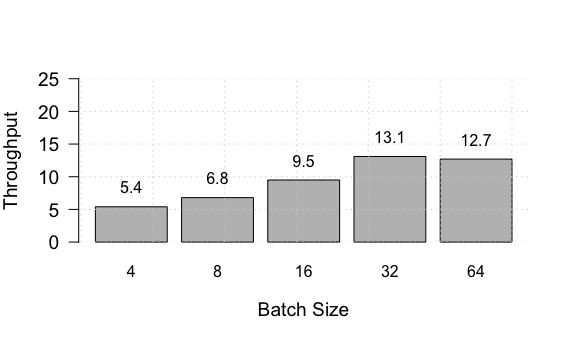}
		\caption{Throughput (in images/sec) -- 1 worker}
		\label{fig:singlenodethroughput1w}
	\end{subfigure} 
	\begin{subfigure}[t]{0.45\textwidth}
		\centering
		\includegraphics[width=\textwidth]{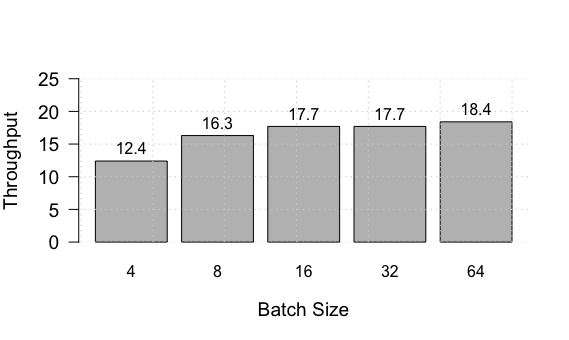}
		\caption{Throughput (in images/sec) -- 4 workers}
		\label{fig:singlenodethroughput4w}
	\end{subfigure}
			\begin{subfigure}[t]{0.45\textwidth}
				\includegraphics[width=\textwidth]{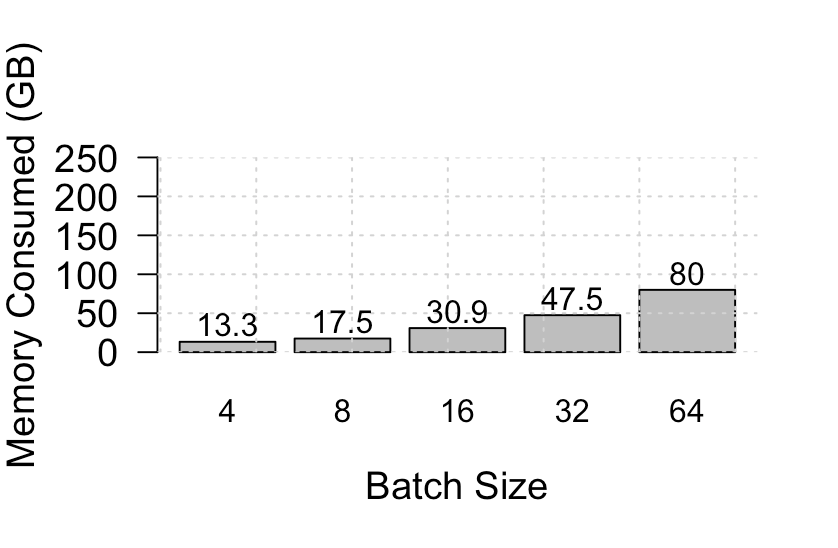}
				\caption{Memory (in GB) -- 1 worker}
				\label{fig:singlenodememutil1w}
			\end{subfigure}
			\begin{subfigure}[t]{0.45\textwidth}
				\includegraphics[width=\textwidth]{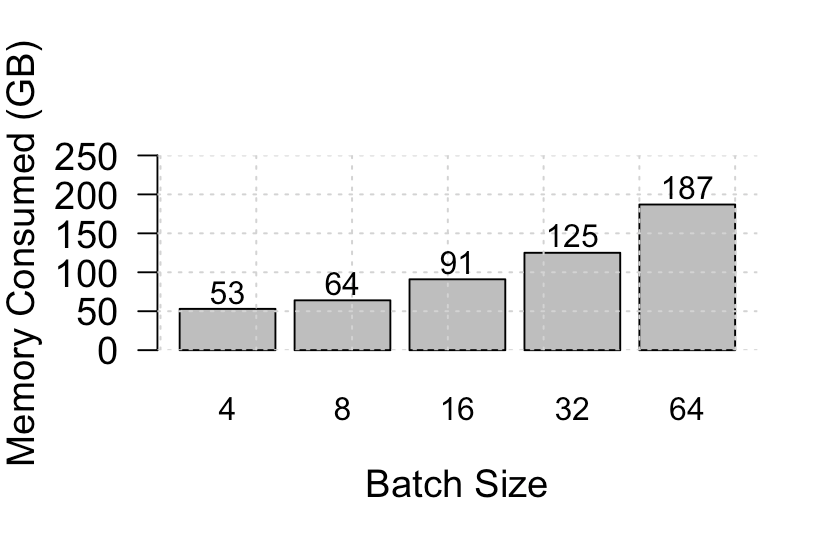}
				\caption{Memory (in GB) -- 4 workers}
				\label{fig:singlenodememutil4w}
			\end{subfigure}
			\caption{\label{fig:multiworker}
				\textsf{Throughput (in images/second) and memory utilized (in GB) with batch sizes 4 to 64 for 1 and 4 training workers respectively (a and b) on a single 2S Intel\textregistered{} Xeon\textregistered{} Gold 6148 processor with Dataset A.}}
		\end{figure}

		\begin{figure}[h!]
			\centering
			\resizebox{0.95\columnwidth}{1.2in}{
				\includegraphics{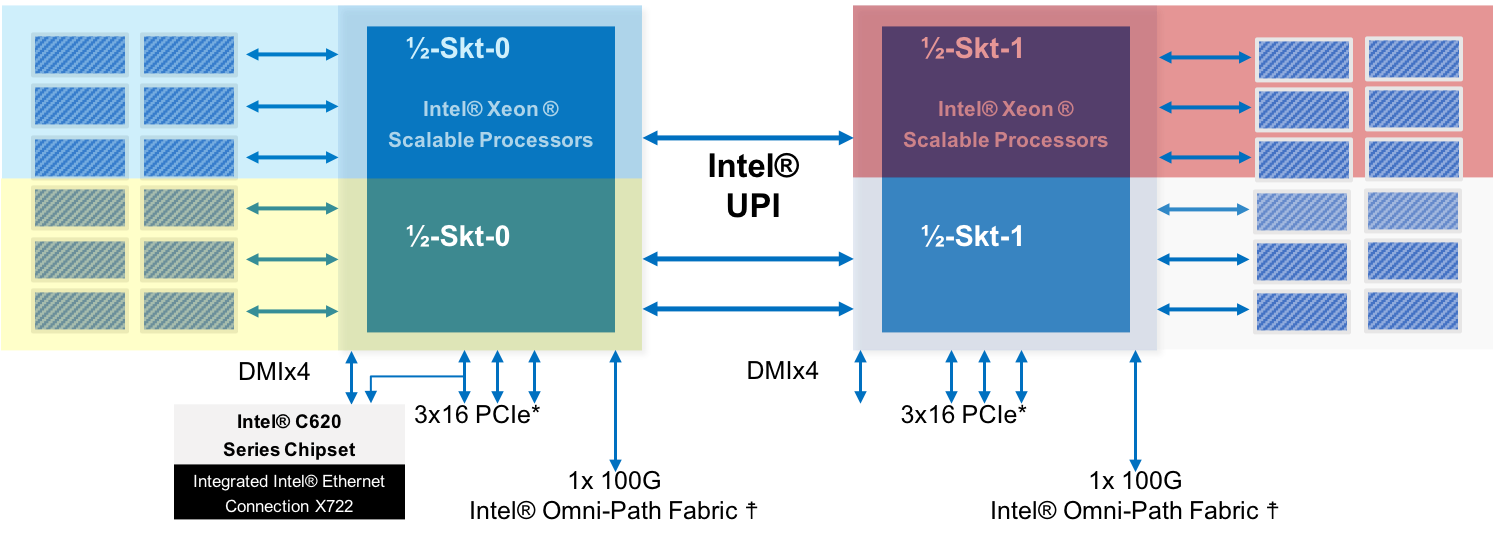}
			}
			\caption{\textsf{Two socket Intel\textregistered{} Xeon\textregistered{} Gold 6148 processor NUMA configuration}}
			\label{fig:multisocket}
		\end{figure}
	
\noindent Second, for all batch size configurations, CPU utilization was low meaning the cores were under-utilized. Upon further investigation with system profile, we found 1) there were lots of context switches and 2) processes or threads assigned to one CPU socket are accessing data from the other CPU socket including a long latency hop over the socket-to-socket interconnect. This led to the discovery that using multiple workers or instanes per socket can yield faster TTT. The essence of using multiple workers in a single CPU is to affinitize tasks to cores and bind their memory allocation to local non-uniform memory access (NUMA) banks as shown by the shaded rectangles in \autoref{fig:multisocket}. Memory binding is key here as it avoids redundant fetches over the interconnect to the memory channels of the adjacent CPU socket. More detailed analysis of multiple workers or instances in training and inference are described in detail by Saletore and colleagues, in \cite{Saletore2018}. \\

\noindent While the authors mention that instantiating multiple workers boosts performance, they do not specify the optimal number of workers, which can depend on a variety of factors, including the neural network topology being trained, CPU micro-architecture, and characteristics of the input data. To find the best combination of workers and local batch size per worker, we experimented with 1, 2, 4 and 8 workers per CPU. In this case, 4 workers with 8 local mini-batch size resulted in the highest throughput per node. A detailed analysis of throughput and memory utilization for 4 workers is shown in Figure \ref{fig:multiworker} \subref{fig:singlenodethroughput4w} and Figure \ref{fig:multiworker} \subref{fig:singlenodememutil4w}, respectively. Note that throughput with batch sizes of 64, 128, or 256 was higher than with a batch size of 32, but these configurations did not converge any faster.
	\end{subsection}
	\begin{subsection}{Scaling out TTT on 8 Servers with Dataset A}
		\label{sec:scaleout_dataset1}
\noindent After determining the number of workers per node, we deployed the training on 8 nodes with 4 workers per node. We used the MPI Allreduce mechanism in Uber's Horovod library to synchronize the gradients. As indicated in \autoref{fig:mcnn}, the model size is 162MB which was the size of the gradients exchanged between the workers per iteration. Due to this high bandwidth requirement, we used a 100Gbps Intel\textregistered{} Omni-Path Fabric (Intel\textregistered{} OP Fabric). Note here that each layer of M-CNN calls \emph{Horovod\_Allreduce}, resulting in a large variation in the MPI negotaition calls. The MPI negotiation times range between 450ms and 858ms. The final time to convergence on 8 nodes is shown in \autoref{fig:8node}. \autoref{fig:8node}\subref{fig:loss8node} shows the training loss over epochs and \autoref{fig:8node}\subref{fig:accuracy8node} shows the time to achieve state of the art top-1 and top-5 accuracy on Dataset A. From the results, we see that using 8x more hardware resources we were able to scale TTT by 6.6X. With Dataset A, this means a TTT of 31 minutes which is well within our target of one hour. This also encouraged us to explore a larger dataset we would need more hardware resources. Hence, we chose Dataset B with 313,282 images. The experiment results follow in the next section.

\begin{figure}
	\begin{subfigure}[b]{0.5\columnwidth}
		\centering
		\includegraphics[height=2.5in]{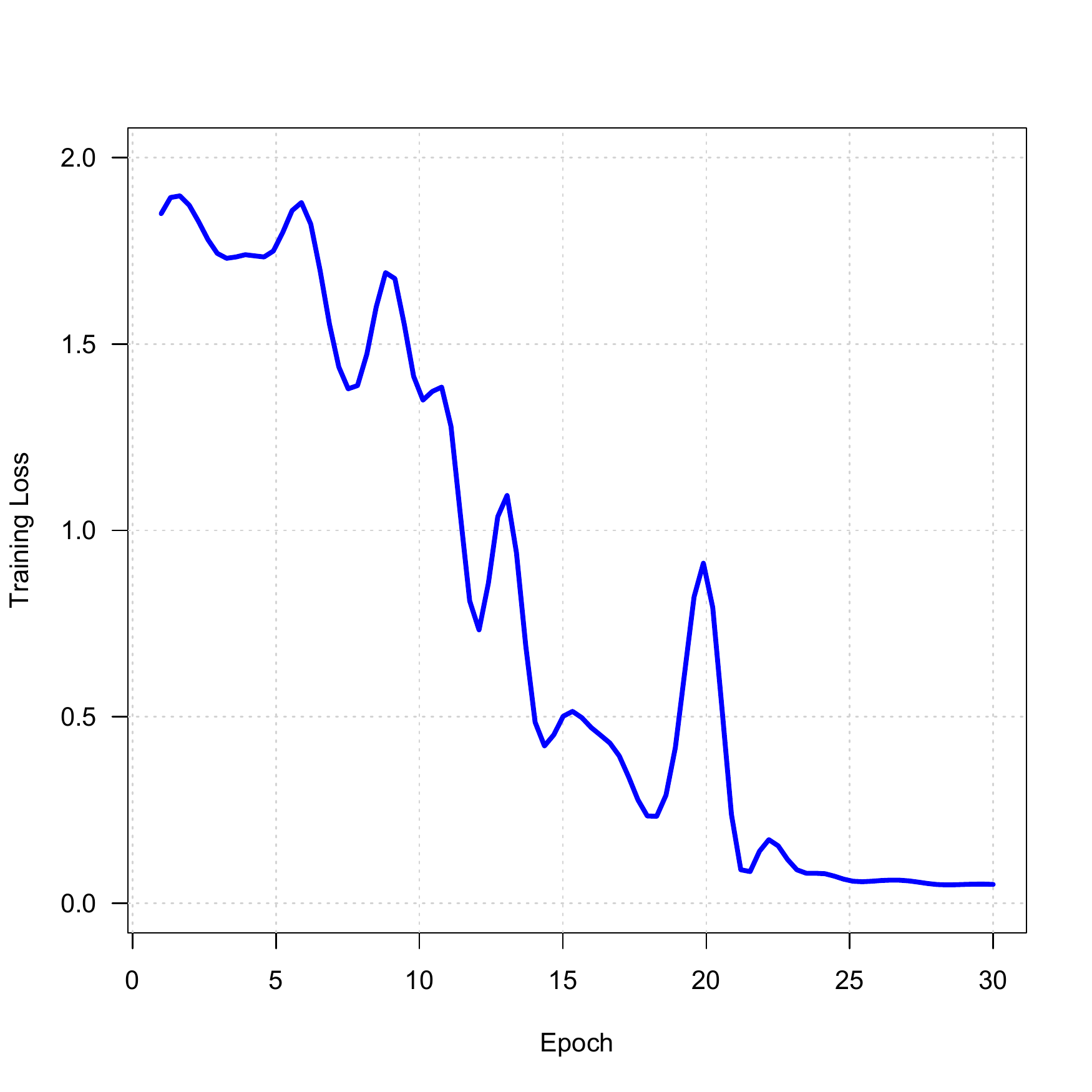}
		\caption{Training loss}
		\label{fig:loss8node}
	\end{subfigure}
	\begin{subfigure}[b]{0.5\columnwidth}
		\centering
		\includegraphics[height=2.5in]{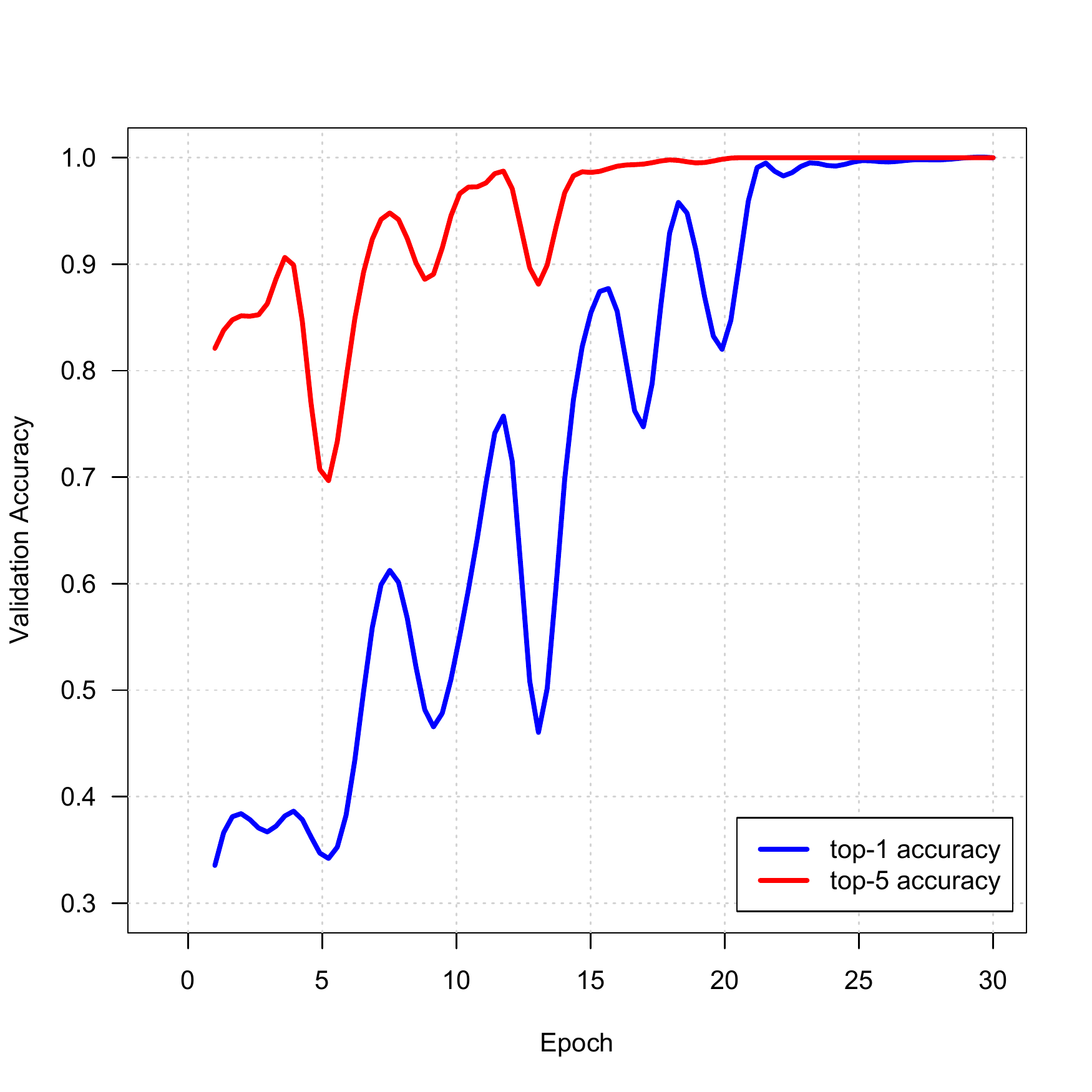}
		\caption{Validation accuracy}
		\label{fig:accuracy8node}		
	\end{subfigure}
	\caption{\label{fig:8node}%
		\textsf{Training loss, top-1 and top-5 accuracy of M-CNN model with Dataset A in 30 epochs on 8x 2S Intel\textregistered{} Xeon\textregistered{} Gold 6148 processors connected with Intel\textregistered{} OP Fabric}}
\end{figure}

\begin{figure}
	\centering
	\includegraphics[width=2.5in]{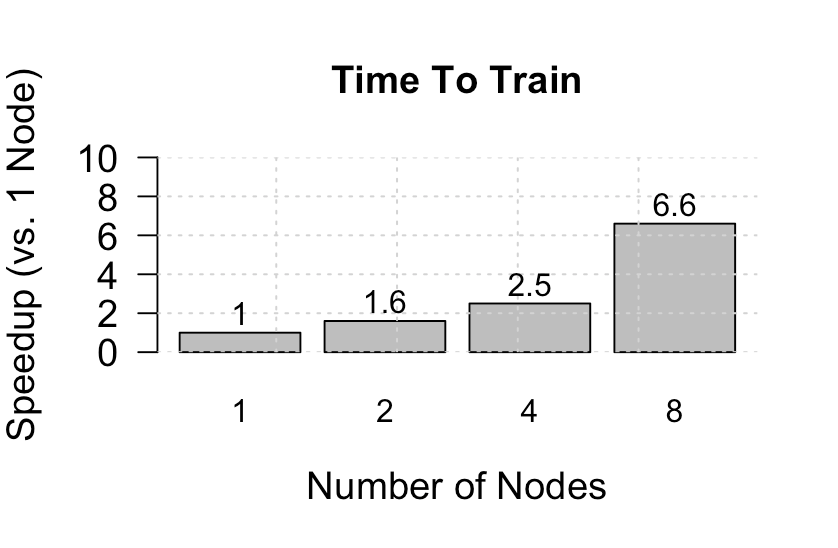}
	\caption{\textsf{Scaling M-CNN training with Dataset A from 1X to 8X 2S Intel\textregistered{} Xeon\textregistered{} Gold 6148 processors connected with 100Gbps Intel\textregistered{} OP Fabric}}
	\label{fig:scaleout_dataset1}
\end{figure}
\end{subsection}

	\begin{subsection}{Scaling out TTT on 128 Servers with Dataset B}
		\label{sec:scaleout_dataset2}
		\begin{table*}[b!]
			\begin{center}
				\caption{M-CNN Training Performance on 128 2S Intel\textregistered{} Xeon\textregistered{} Gold processors with Dataset B}
				{\renewcommand{\arraystretch}{1.5}%
					\begin{tabular}{|c|c|c|c|c|}
						\hline
						\textbf{\# of Nodes} & \textbf{\# of Epochs} & \textbf{Batch Size} & \textbf{TTT (mins)} & \textbf{Images/sec}\\
						\hline
						\hline
						1 & \centering6.6 & 128 & 960 & 30\\
						2 & \centering8 & 256 & 642 & 72\\
						4 & \centering8.7 & 512 & 320 & 141\\
						8 & \centering12 & 1024 & 240 & 262\\
						16 & \centering15.9 & 2048 & 150 & 553\\
						32 & \centering14.9 & 2048 & 85 & 893\\
						64 & \centering15 & 2048 & 61 & 1284\\
						128 & \centering15.2 & 2048 & 50 & 1587\\
						\hline
				\end{tabular}}
				\label{table:mcnn128nodes} 
			\end{center}
		\end{table*}
	
\noindent \autoref{table:mcnn128nodes} summarizes the 19.2X performance improvement acheived by scaling from 1 to 128 Intel\textregistered{} Xeon\textregistered{} Gold 6148 processors with Dataset B bringing TTT to 50 minutes. The second column in the table shows number of epochs when training reach 99\% top-1 accuracy and 100\% top-5 accuracy. Subsequent columns show the global mini-batch size, time to train (in minutes) and effective throughput in images/second for each node configuration. 8 training workers per node were used in these experiments as the image dimensions in Dataset B are smaller than Dataset A.\\

\noindent The key takeaway here is that updates per epoch is critical to acheive convergence. Global mini batch size determines the number of updates per epoch and M-CNN did not converge beyond global batch sizes of 2048. Hence, we maintained the global batch size to 2048 while scaling from 16 to 128 nodes -- the idea of strong scaling taken from HPC applications. As the same amount of work is increasingly divided across more CPU cores we observe diminishing returns in speedup albeit overall TTT improves. Note that our objective is to not show linear scaling here, but to see what resources will help us acheive a TTT less than one hour. \\\\
\noindent Anothe key takeaway is that large number of workers required larger number of epochs to converge. This also affects scaling. This is again an artifact of the dataset. Finally, in \autoref{fig:top1acc_dataset2}, we show the behavior of top-1 accuracy and learning rate per epoch for each of the configurations. Note here that use the linear learning rate scaling rule discussed in \autoref{sec:lr}. The learning rate is scaled according to the ratio of increase in global mini batch size. However, as shown in the \autoref{fig:top1acc_dataset2} similar to global batch size, learning rate scaling has to capped to 2048 beyond 16 nodes for the model to converge. \\\\
		\begin{figure*}[t!]
			\centering
			\begin{subfigure}[b]{0.25\textwidth}
				\includegraphics[width=1.5in]{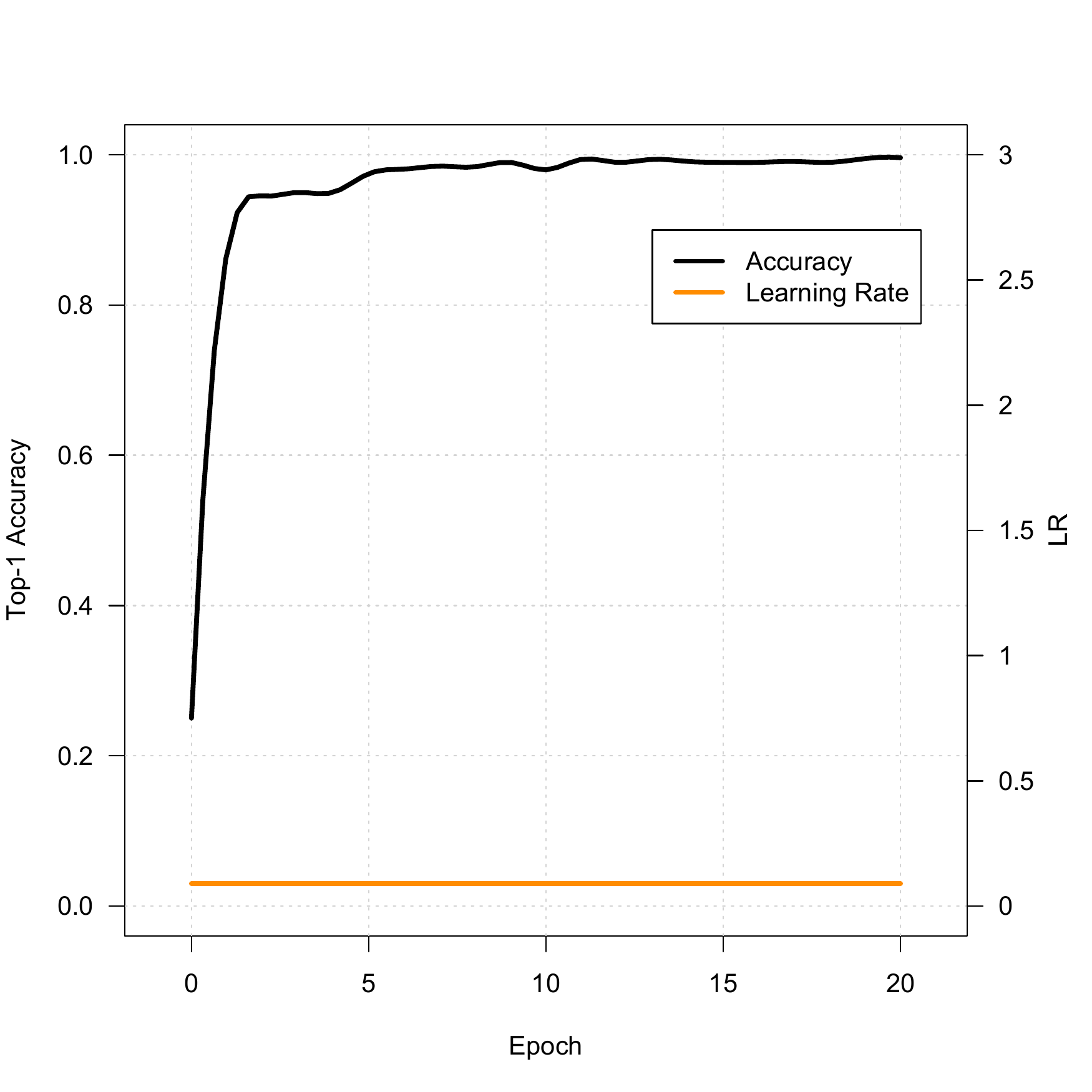}
			\caption{1 node}
			\end{subfigure}%
			\begin{subfigure}[b]{0.25\textwidth}
				\includegraphics[height=1.5in]{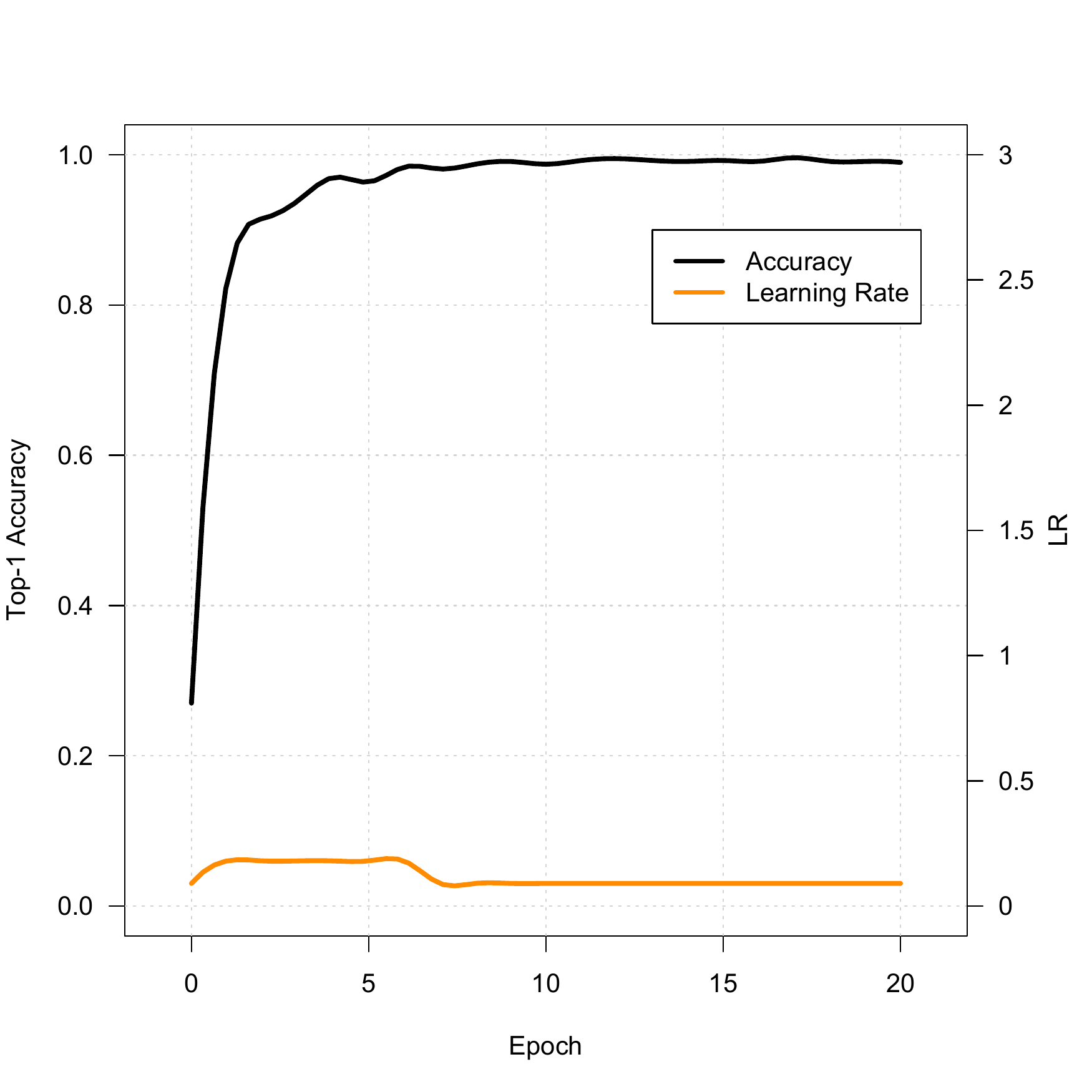}
			\caption{2 nodes}
			\end{subfigure}%
		    \begin{subfigure}[b]{0.25\textwidth}
		   		\includegraphics[height=1.5in]{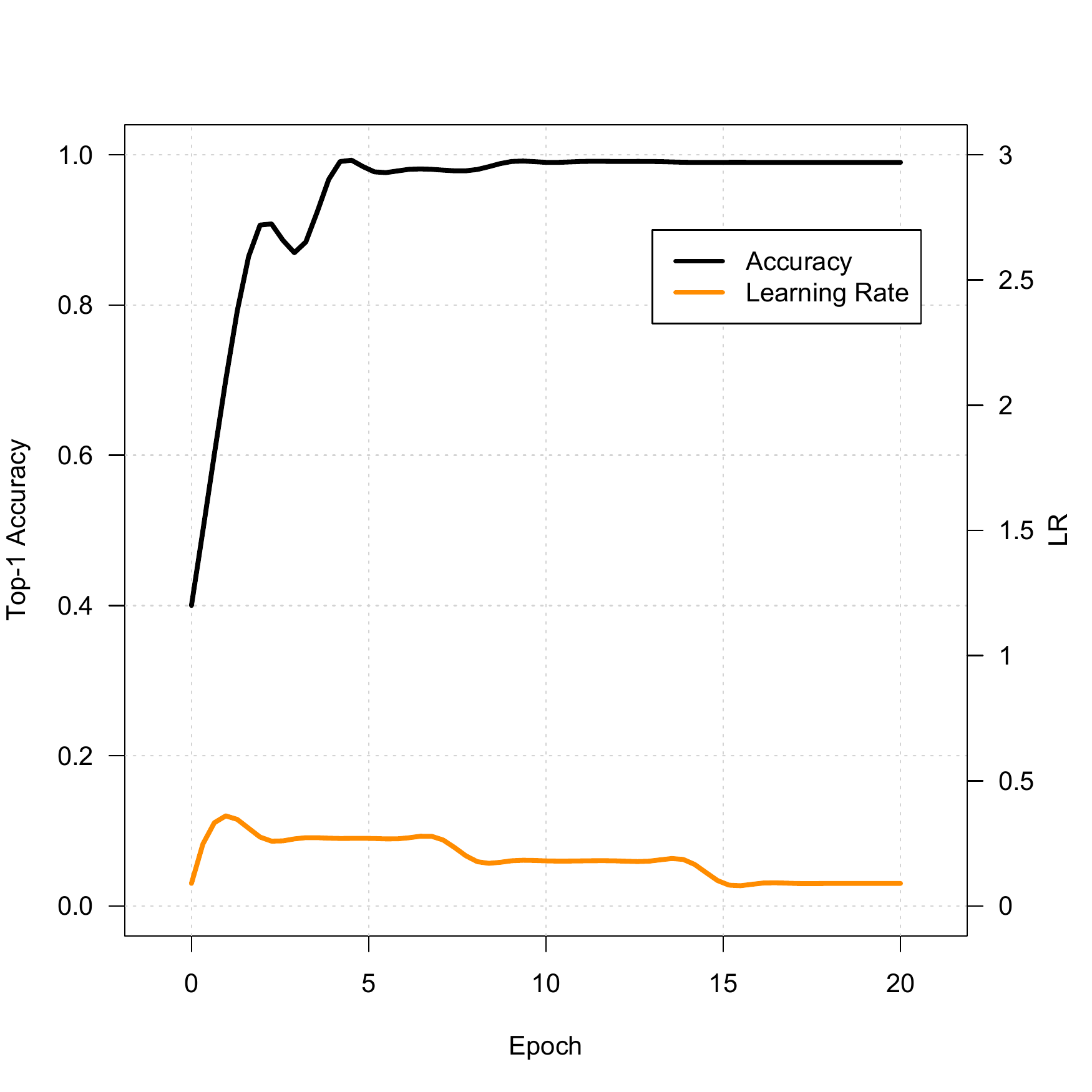}
		    \caption{4 nodes}
		    \end{subfigure}%
		    \begin{subfigure}[b]{0.25\textwidth}
		   		\includegraphics[height=1.5in]{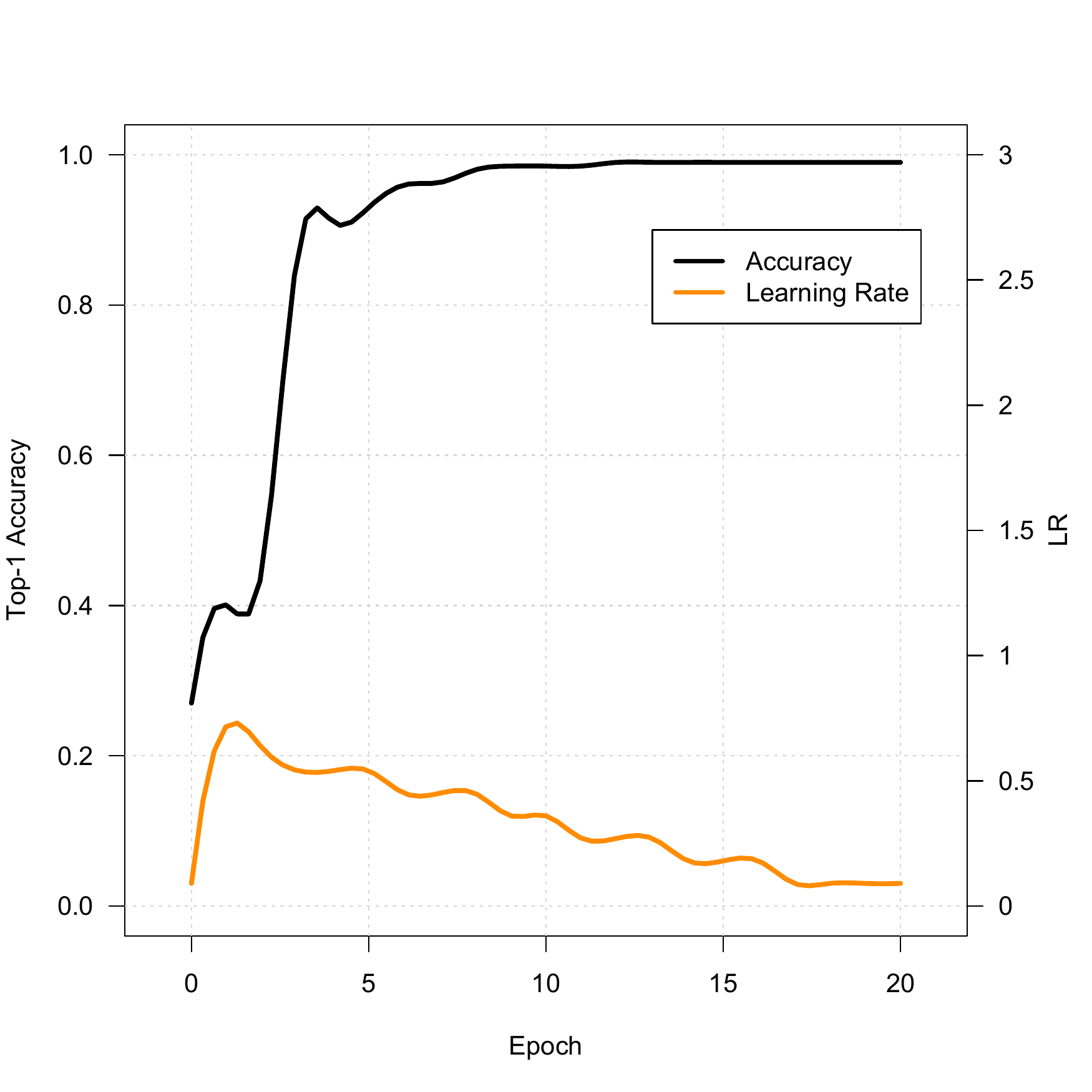}
		    \caption{8 nodes}
		    \end{subfigure}
		    \begin{subfigure}[b]{0.25\textwidth}
		    	\includegraphics[height=1.5in]{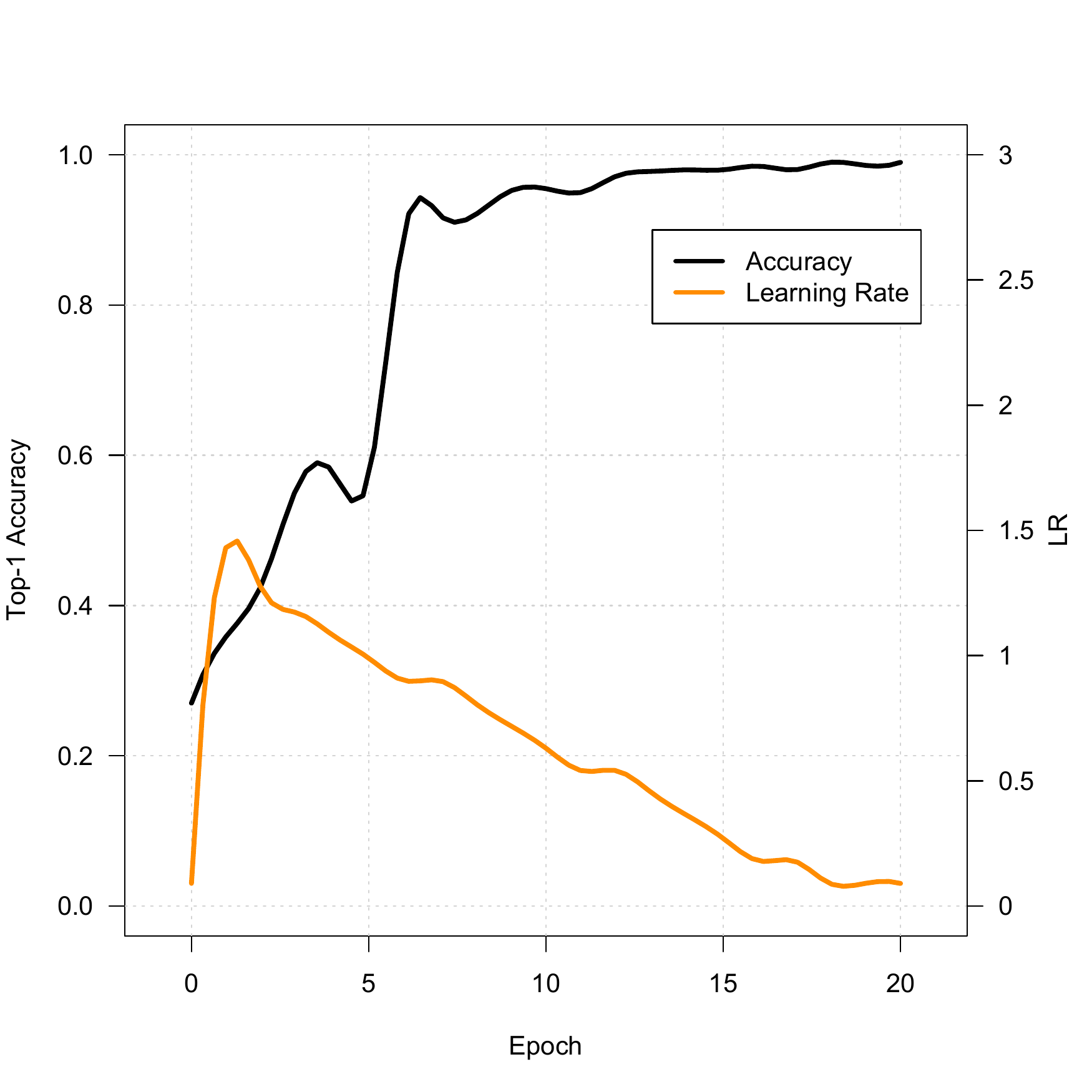}
		    \caption{16 nodes}
		    \end{subfigure}%
		    \begin{subfigure}[b]{0.25\textwidth}
		    	\includegraphics[height=1.5in]{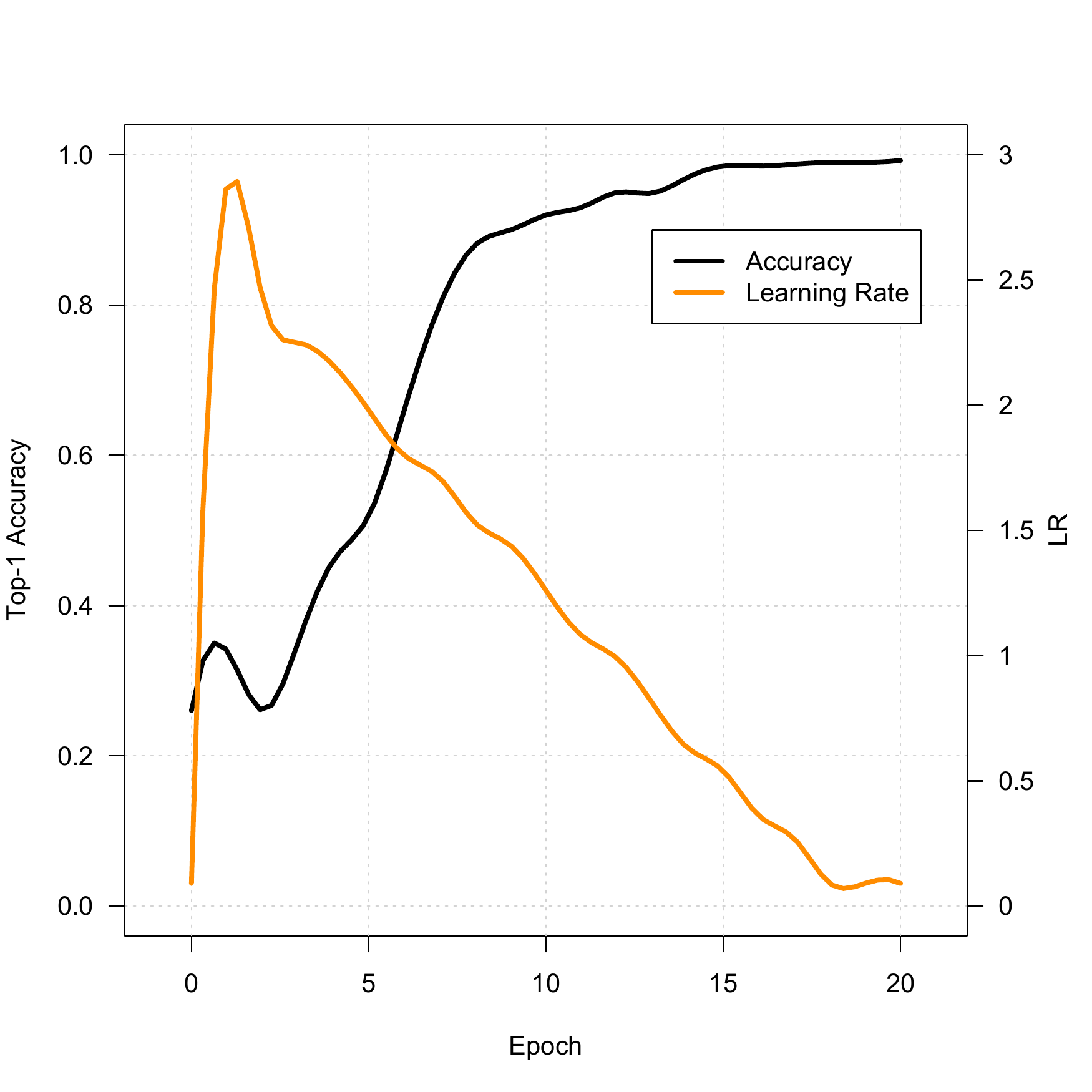}
		    \caption{32 nodes}
		    \end{subfigure}%
			\begin{subfigure}[b]{0.25\textwidth}
				\includegraphics[height=1.5in]{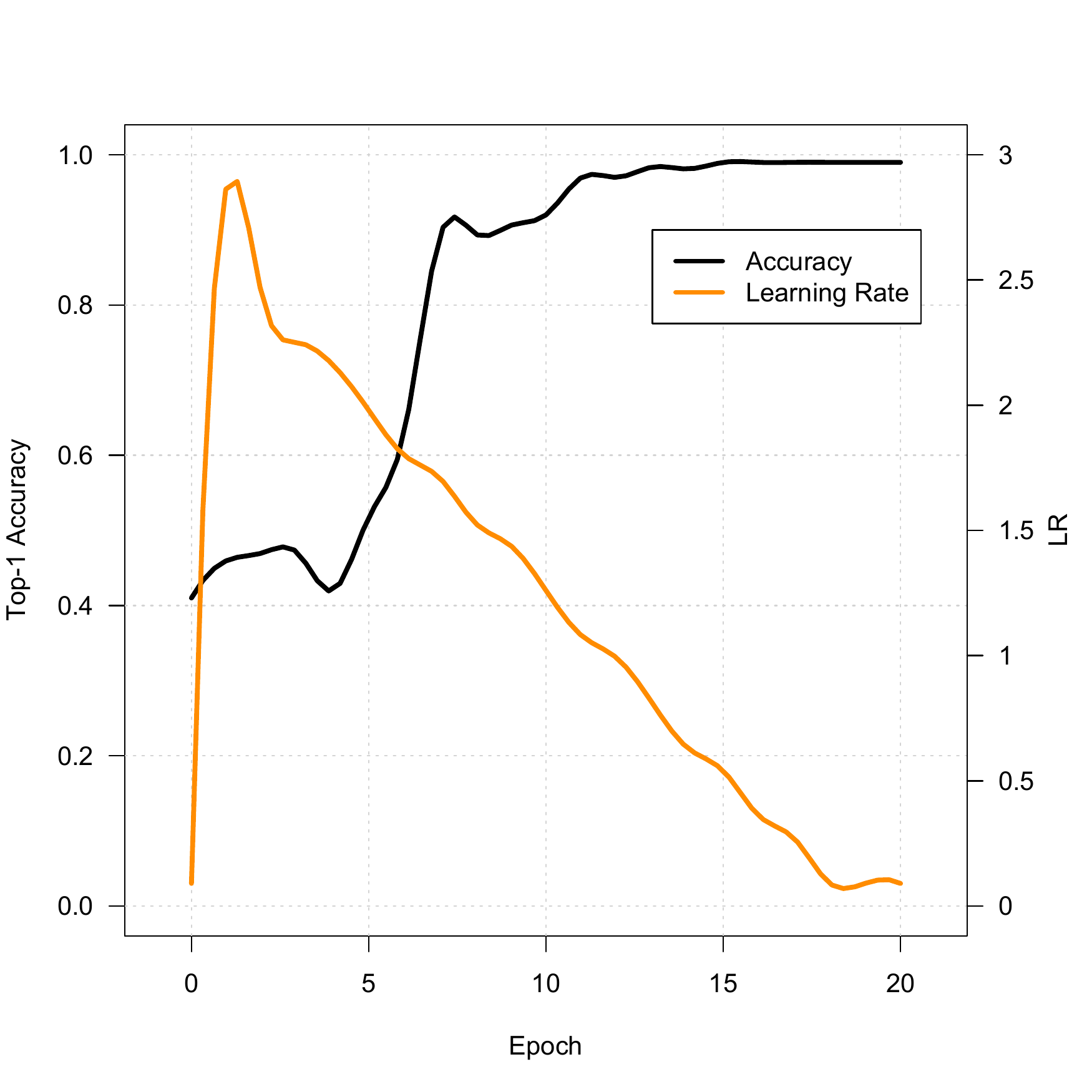}
			\caption{64 nodes}
			\end{subfigure}
			\caption{Top-1 Accuracy achieved in 20 epochs of M-CNN training and Learning Rate used on 1--64 2S Intel\textregistered{} Xeon\textregistered{} Gold processors. Dataset B is used for these experiments. Global minibatch size is capped at 2K from 16 to 64 nodes. The learning rate as shown in (f) -- (h) is also scaled only to 0.032 to achieve convergence}
			\label{fig:top1acc_dataset2}
		\end{figure*}
		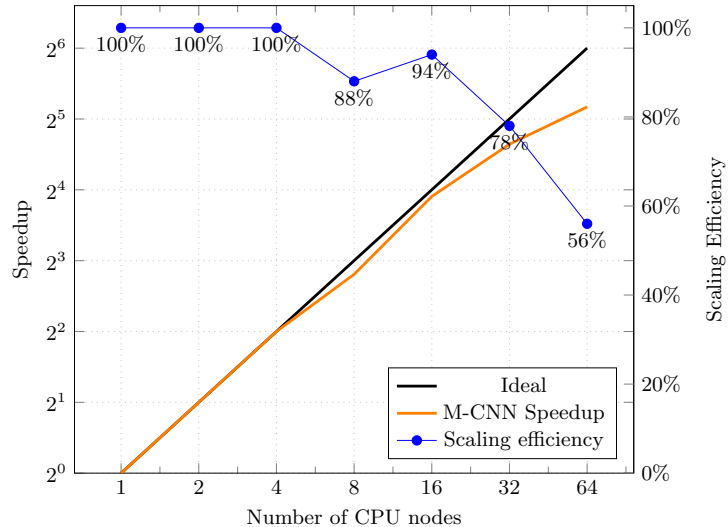
\begin{figure*}
			\centering
			\resizebox{0.8\columnwidth}{!}{
				\begin{tikzpicture}
				\begin{axis}[minor tick num=1,
				xlabel=Number of CPU nodes,
				xtick=data,
				symbolic x coords={1,2,4,8,16,32,64,128},
				ymin=1,
				ylabel=Speedup,
				ymode=log,
				log basis y={2},
				grid=major,
				major grid style={dotted}]
				\addplot [black,very thick] coordinates {(1,1) (2,2) (4,4) (8,8) (16,16) (32,32) (64,64)}; 
				\label{plotOne}
				\addplot [orange,very thick] coordinates {(1,1) (2,2) (4,4) (8,7) (16,15) (32,25) (64,36)}; 
				\label{plotTwo}
				\end{axis}
		        
				\begin{axis}[
				axis y line*=right,
				axis x line=none,
				xtick=data,
				symbolic x coords={1,2,4,8,16,32,64}, 
				ymin=0, ymax=105,
				ylabel=Scaling Efficiency,
				ytick=\empty,
				extra y ticks={0,20,40,60,80,100},
				extra y tick labels={0\%,20\%,40\%,60\%,80\%,100\%},
				nodes near coords={\pgfmathprintnumber\pgfplotspointmeta\%},
				nodes near coords align=below,
				every node near coord/.append style={color=black},
				legend pos=south east]
				\addlegendimage{/pgfplots/refstyle=plotOne}\addlegendentry{Ideal}
				\addlegendimage{/pgfplots/refstyle=plotTwo}\addlegendentry{M-CNN Speedup}
				\addplot [mark=*,blue] coordinates {(1,100) (2,100) (4,100) (8,88) (16,94) (32,78) (64,56)}; 
				\addlegendentry{Scaling efficiency}
				\end{axis}
				\end{tikzpicture}}
			\caption{Scalability of M-CNN training performance for 20 epochs on 64 2S Intel\textregistered{} Xeon\textregistered{} Gold 6148 processors. Note that global batch size is capped at 2K from 16 -- 64 nodes. Intel\textregistered{} OP Fabric, TensorFlow-1.9.0+Horovod, OpenMPI v3.0.0, 8 workers/node}
			\label{fig:20epoch_dataset2}
		\end{figure*}
\noindent Additionally, we show the scaling efficiency of M-CNN training from 1 to 64 nodes all running for 20 epochs. As shown in \autoref{fig:20epoch_dataset2} time to train efficiently scales up to 16 nodes after which capping the global mini batch size shows diminishing returns.
	\end{subsection}
\end{section}

%% file: discussion.tex
\begin{section}{Discussion} \label{sec:discussion}
	\noindent  In this work, we explored training a multi-scale convolutional neural network to classify large high content screening images within one hour by exploiting large memory in CPU systems. The cellular images used are over million pixels in resolution and are 26 times larger than those in the ImageNet dataset. We used two sets of cellular images with different resolutions to analyze the performance on multiple nodes of M-CNN training. The first set contains 10K full resolution cellular images (1024$\times$1280$\times$3) and the second dataset contains 313K images of smaller dimensions (724$\times$724$\times$3). With the first dataset, we were able to scale time to train linearly using 8X 2S Intel\textregistered{} Xeon\textregistered{} Gold processors. Large mini-batch sizes enabled by the large memory footprint in CPUs helped us achieve the speedup in training time. With the second data set, we were able to achieve TTT of 50 minutes, a 19.2X improvement in time to train using 128 Intel\textregistered{} Xeon\textregistered{} Gold processors. We learned that the updates per epoch is critical to achieve convergence and if the characteristics of the images in the dataset cannot tolerate scaling of updates per epoch beyond a certain threshold (2048 in our case), then adding more computational resources results in diminishing returns. In future work, we intend to explore larger datasets with more variation where images are chosen from different cohorts. \\

  \noindent \textbf{Acknowledgements.} We would like to acknowledge Wolfgang Zipfel from the Novartis Institutes for Biomedical Research, Basel, Switzerland; Michael Derby, Michael Steeves and Steve Litster from the Novartis Institutes for Biomedical Research, Cambridge, MA, USA; Deepthi Karkada, Vivek Menon, Kristina Kermanshahche, Mike Demshki, Patrick Messmer, Andy Bartley, Bruno Riva and Hema Chamraj from Intel Corporation, USA, for their contributions to this work. The authors also acknowledge the Texas Advanced Computing Center (TACC) at The University of Texas at Austin for providing HPC resources that have contributed to the research results reported within this paper. URL: http://www.tacc.utexas.edu. \\

  \noindent
	\textbf{Conflicts of interest.} Intel\textregistered{} Xeon\textregistered{} Gold 6148 processor, Intel\textregistered{} OPA and Intel\textregistered{} SSD storage drive are registered products of Intel Corporation. The authors declare no other conflicts of interest.

\end{section}